\documentclass[twocolumn,aps,prc,superscriptaddress,showpacs,floatfix]{revtex4}
\usepackage{amssymb}
\usepackage{amsmath,bm}
\usepackage[normalem]{ulem}
\usepackage[dvips]{color}
\usepackage{booktabs}
\usepackage{threeparttable}
\usepackage{graphicx}
\usepackage{CJK}
\renewcommand{\sout}{\bgroup \color[rgb]{1,0,0}\ULdepth=-.5ex \ULset}

\begin{document}

\title{Analytical coalescence formula for particle production in relativistic heavy-ion collisions}

\author{Kai-Jia Sun}
\affiliation{Department of Physics and Astronomy and Shanghai Key Laboratory for
Particle Physics and Cosmology, Shanghai Jiao Tong University, Shanghai 200240, China}
\author{Lie-Wen Chen\footnote{%
Corresponding author (email: lwchen$@$sjtu.edu.cn)}}
\affiliation{Department of Physics and Astronomy and Shanghai Key Laboratory for
Particle Physics and Cosmology, Shanghai Jiao Tong University, Shanghai 200240, China}
\date{\today}

\begin{abstract}
Based on a covariant coalescence model with a blast-wave-like parametrization for the
phase-space configuration of constituent particles at freeze-out, we derive an approximate
analytical formula for the yields of clusters produced in relativistic heavy-ion
collisions.
Compared to previous existing formulae, the present work additionally considers the contributions from the longitudinal
dimension in momentum space, the relativistic corrections and the finite size effects of
the produced clusters relative to the spatial distribution of constituent particles
at freeze-out.
The new analytical coalescence formula provides a useful tool to evaluate the yield
of produced clusters, such as light nuclei from nucleon coalescence and hadrons from
quark coalescence, in heavy-ion collisions.
As a first application of the new analytical formula, we explore the strangeness population factor
$S_3 = ^3_{\Lambda}$H/($^3$He$\times \Lambda/$p) based on nucleon/$\Lambda$ coalescence
as well as the production of exotic hadrons based on quark coalescence,
in central Pb+Pb collisions at $\sqrt{s_{NN}}=2.76$~TeV.
The results are compared with the predictions from other models.

\end{abstract}

\maketitle

\section{Introduction}
\label{introduction}
Particle production in relativistic heavy-ion collisions is of fundamental importance
for many issues in nuclear physics, particle physics, astrophysics and cosmology.
Coalescence model~\cite{But61,Sat81,Cse86,Dov91} provides an important approach to
describe the particle production in heavy-ion collisions at both intermediate and
high energies.
For instance, the coalescence model has been successfully and widely applied to
describe both light nuclei production from nucleon coalescence~\cite{Gyu83,Mat97,ChenLW03,Oh09,Ste12,ChenG12,ChenG13,Zhu15,Zha10,Xue15,KJSUN15,KJSUN16,KJSUN16-1}
and hadron production from quark coalescence in heavy-ion
collisions~\cite{Lin02,Vol03,Hwa03,Gre03,Fri03,Mol03,Pra05,Sha05,LWChen06,Ko14,Fri08}.
The quark coalescence provides an important mechanism for the hadronization of
partons produced in relativistic heavy-ion collisions and thus is very useful for
understanding the partonic dynamics as well as the formation signals and properties
of quark-gluon plasma possibly formed in these collisions~\cite{Fri08,Adl03,Ada04}.
Since the coalescence probability is based on the overlap of density matrix of
constituent particles in an emission source with Wigner function of the produced
cluster, the predicted cluster yield generally depends on the
internal structure wave function of the cluster.
Therefore, the coalescence model provides a unique tool to
identify the constituent quark structure of some exotic hadrons via quark coalescence
from studying their yields in relativistic heavy-ion
collisions~\cite{ChenLW04,ChenLW07,Cho11,Cho11prc}.

To describe particle production in heavy-ion collisions, the coalescence model
is usually implemented with the phase-space configuration of constituent particles
at freeze-out in an emission source obtained from either transport model
simulations~\cite{Gyu83,Mat97,ChenLW03,Oh09,Ste12,ChenG12,ChenG13,Zhu15,LWChen06,Ko14} or
empirical parametrization~\cite{Gre03,Fri03,ChenLW04,ChenLW07,Zha10,Xue15,KJSUN15,KJSUN16,KJSUN16-1},
and the final results are then obtained by multi-dimension numerical integration.
Although the numerical calculations
can give exact results in the coalescence model, the approximate analytical formula
is extremely useful and has its own merits,
e.g., it can significantly simplify the computational work and most importantly it can give
more transparent and deep insights into the physics of the coalescence scenario.

In the literature,
there indeed exist some analytical coalescence
formulae (see, e.g.,~\cite{ChenLW07,Cho11,Cho11prc}) for cluster yields in relativistic
heavy-ion collisions, obtained under some approximations. In particular, the analytical formula
reported recently by the ExHIC collaboration in Refs.~\cite{Cho11,Cho11prc} (denoted as
COAL-Ex in the present work) is quite general for $N$-body cluster production with the $N$
constituent particles at various orbit angular momentum states.
In the derivation of the COAL-Ex formula, for simplicity, the
integration of Wigner function in longitudinal ($z$-) direction is neglected, namely,
the integration in momentum space is only treated in two-dimensional transverse plane 
although the integration in coordinate space is treated three-dimensionally, which leads to a
problematic feature that the yield of the produced cluster in the COAL-Ex formula increases
as a power function of its root-mean-square (RMS) radius $r_{\text{rms}}$
when $r_{\text{rms}}$ becomes large.
Even though the single-particle momentum distribution for a particle in the emission source
may only depend on its transverse momentum due to boost invariance~\cite{Bjo83} in the longitudinal
direction, the many-body coalescence process cannot be treated merely in two-dimensional
transverse plane.
It is thus of great interest to include the contribution of longitudinal momentum dimension
and check its importance.

In the present work, based on a blast-wave-like parametrization, which is inherently
longitudinal boost invariant by design, for the emission source of constituent particles at freeze-out,
we derive a new analytical coalescence formula (denoted as COAL-SH in the following)
by treating the integration of Wigner function in full
phase-space for the many-body coalescence process.
We find that the COAL-SH formula possesses a nice saturation property that when $r_{\text{rms}}$
of the produced cluster is large, its yield converges to a constant.
In addition, we consider the relativistic corrections to leading order in the COAL-SH formula.
Furthermore, an empirical expression is proposed for the corrections of
finite size effects of the produced clusters relative to the size of the emission source
for constituent particles.
As a first application of the new analytical coalescence formula,
we explore the strangeness population factor $S_3 = ^3_{\Lambda}$H/($^3$He$\times \Lambda/$p)
as well as the production of exotic hadrons in central Pb+Pb collisions at
$\sqrt{s_{NN}}=2.76$~TeV, and compare the results with the predictions from other models.

\section{Theoretical formalism}
\label{formulism}

\subsection{Blast-wave-like parametrization for emission source of constituent particles at freeze-out}
We assume that particles are emitted from a hypersurface $\Sigma^\mu$, and then the Lorentz invariant
one-particle momentum distribution is given by
\begin{eqnarray}
\label{EqA1}
E\frac{\text{d}N}{\text{d}^3p} = \int\limits_{\Sigma^\mu}\text{d}\sigma _{\mu} p^\mu f(x,p) = \int \text{d}^4x S(x,p),
\end{eqnarray}
where $\sigma_\mu$ denotes the normal vector of the hypersurface $\Sigma^\mu$ and $p^\mu$ is the
four-momentum of the emitted constituent particle.
For the cluster production at midrapidity in relativistic heavy-ion collisions
that we are considering in this work, we adopt the longitudinal boost invariance
assumption~\cite{Bjo83} and assume the constituent particles are emitted at a fixed proper time $\tau_0$,
and the emission function can then be expressed as~\cite{Ret04}
\begin{eqnarray}
\label{EqA2}
S(x,p)\text{d}^4x &=&  m_T\cosh(\eta-y)f(x,p)    \notag \\
     && \times \delta(\tau-\tau_0) \tau \text{d}\tau\text{d}\eta r \text{d}r \text{d}\phi_s,
\end{eqnarray}
where we use the longitudinal proper time $\tau = \sqrt{t^2-z^2}$, space-time rapidity
$\eta = \frac{1}{2} \text{ln}\frac{t+z}{t-z}$, polar coordinates ($r$, $\phi_s$),
rapidity $y=\frac{1}{2}\ln (\frac{E+p_z}{E-p_z})$, transverse momentum ($p_T,\phi_p$)
and transverse mass $m_T=\sqrt{m^2+p_T^2}$.
The statistical distribution function $f(x,p)$ is given
by~\cite{Coo74} $f(x,p)=g(2\pi)^{-3}[\exp(p^{\mu}u_{\mu}/kT)/\xi \pm 1]^{-1}$
where $g$ is spin degeneracy factor, $\xi$ is the fugacity, $u_{\mu}$ is the
four-velocity of a fluid element in the fireball of the emission source and $T$ is the
corresponding local temperature. In addition, we note
$p^{\mu} u_{\mu}=m_T \cosh\rho \cosh(\eta -y)-p_T \sinh\rho \cos(\phi_p -\phi_s)$ gives
the energy in local rest frame of the fluid, and
$ p^{\mu} \text{d}^3\sigma_{\mu}=\tau m_T \cosh(\eta -y)\text{d}\eta r\text{d}r\text{d}\phi_s$.
The symbol $\rho$ represents the transverse flow rapidity distribution of the fluid
element in the fireball with a transverse radius $R$.

Assuming the temperature of fireball is much smaller than the mass of constituent particles,
we can then use the Boltzmann approximation for $f(x,p)$, and Eq.~(\ref{EqA1}) can then be
analytically obtained as~\cite{Ret04}
\begin{eqnarray}
\label{EqA3}
\frac{\text{d}^3N}{p_T\text{d}p_T\text{d}y\text{d}\phi_p} = \frac{g\xi\tau m_T}{(2 \pi)^3} \int 2K_1(\beta) 2\pi I_0(\alpha) r dr,
\end{eqnarray}
where we have $\beta = m_T\cosh[\rho(r)] /T$ and $\alpha = p_T \sinh[\rho(r)]/T$;
$I_0(x)$ and $K_1(x)$ are the first and second kind of modified Bessel functions, respectively.
The radial expansion of the fireball would lead to a blue-shift
on the transverse spectrum of the emitted particles, and thus effectively increase the temperature.
As a result, we can approximately take $T$ as an effective temperature and set radial flow
rapidity $\rho$ to be zero, i.e., $\alpha = 0$ and $I_0(0) = 1$, and in this case
the above formula can be further simplified to be
\begin{eqnarray}
\label{EqA4}
\frac{\text{d}^3N}{p_T\text{d}p_T\text{d}y\text{d}\phi_p} &=& \frac{g\xi V}{(2\pi)^3} 2m_T K_1\Big(\frac{m_T}{T}\Big),
\end{eqnarray}
where we denote $V = \pi R^2\tau$ as an effective volume. Furthermore, the multiplicity of
the constituent particles can be integrated out to be
\begin{eqnarray}
\label{EqA5}
\frac{dN}{dy}
&=& \frac{g\xi V}{(2\pi)^3} 4\pi Tm^2 K_2\Big(\frac{m}{T}\Big).
\end{eqnarray}

In the case of $m \gg T$, according to the asymptotic behavior of Bessel
function $K_{\nu} (x)$ in large $x$ limit, i.e.,
\begin{eqnarray}
\label{EqA5-1}
 K_{\nu} (x) \rightarrow \sqrt{\frac{\pi}{2x}} e^{-x} (1+\frac{4\nu^2 - 1}{8x}+\mathcal{O}(\frac{1}{x^2})),
\end{eqnarray}
we can keep only the first term to make a non-relativistic approximation, and then
Eq. (\ref{EqA5}) can be written as
\begin{eqnarray}
\label{EqA6}
\frac{dN}{dy} &=& \frac{g}{(2\pi)^3}\xi \text{e}^{-\frac{m}{T}} VV_{p},
\end{eqnarray}
where $V_{p} = (2\pi T m)^{\frac{3}{2}}$ reflects the effective volume of
the emission source in momentum space.
If we denote $y_L = y_{\text{max}} - y_{\text{min}}$, $V' = y_L V$ and
$\xi' = \xi e^{-\frac{m}{T}}$, then the number of the emitted constituent particle in
rapidity region [$y_{\text{min}},~y_{\text{max}}$] is given by
\begin{eqnarray}
\label{EqA7}
N = \frac{g}{(2\pi)^3} \xi' V' V_{p}.
\end{eqnarray}

\subsection{Analytical coalescence formula}
\label{AnaCoalFormula}
We consider the case that $N$ constituent particles are coalesced into a
cluster, and the total multiplicity of the cluster can be expressed as
\begin{eqnarray} \label{EqB1}
N_c&=&g_c\int \Big(\prod_{i=1}^{N}\text{d}N_i\Big)\rho_c^W(x_1,...,x_N;p_1,...,p_N) \notag \\
&=&g_c\int \bigg(\prod_{i=1}^{N}   p_{i}^\mu\text{d}\sigma_{i\mu}\frac{\text{d}^3p_i}{E_i}f(x_i,p_i)\bigg)\times \notag \\
   &&\rho_c^W(x_1,...,x_N;p_1,...,p_N),
\end{eqnarray}
where $\rho_c^W(x_1,...,x_N;p_1,...,p_N)$ is the Wigner density function which
gives the coalescence probability, and $g_c$ is the coalescence factor~\cite{Sat81}.
Eq.~(\ref{EqA7}) can be used to normalize the coalescence probability and so
Eq.~(\ref{EqB1}) can be rewritten as
\begin{eqnarray} \label{EqB2}
N_c &=&g_c \bigg[\prod_{i=1}^{N} N_i \bigg] \frac{ \int
   \rho_c^W \prod_{i=1}^{N}   p_{i}^\mu\text{d}^3\sigma_{i\mu}\frac{\text{d}^3p_i}{E_i}f(x_i,p_i) }{\prod_{i=1}^{N} \frac{g_i}{(2\pi)^3} \xi'_i V' V_{p,i}   }, \notag  \\
\end{eqnarray}
where $N_i = \frac{g_i}{(2\pi)^3} \xi_i' V' V_{p,i}$ is the multiplicity of
the $i$-th constituent particle.
The Wigner function in the above formula cannot be analytically integrated
out and the result is usually obtained by a multi-dimension numerical
integration method (see, e.g., Ref.~\cite{KJSUN15}).

Following Refs.~\cite{Cho11,Cho11prc}, we consider the produced clusters are in
midrapidity region where one has $\tau \simeq t$, and the volume element on the
hypersurface $p_i^\mu d\sigma_{i\mu} \text{d}^3p_i/E_i$ can then be approximated
to be $\text{d}^3x_i\text{d}^3p_i$.
Furthermore, the non-relativistic Boltzmann approximation
for the distribution function $f(x,p)$ is adopted in the following derivation, and thus one has
$V_{p,i}=(2\pi Tm_i)^{\frac{3}{2}}$. We will
discuss the corrections of relativistic effects later.
Therefore, Eq.~(\ref{EqB2}) can be approximated as
\begin{eqnarray} \label{EqB3}
N_c   &\approx & g_c \bigg[\prod_{i=1}^{N} N_i \bigg] \frac{ \int
   \rho_c^W \prod_{i=1}^{N}   \text{e}^{-\frac{p_{T,i}^2}{2m_i T}\cosh(\eta_i - y_i)}\text{d}^3x_i \text{d}^3p_i }{\prod_{i=1}^{N}  V' V_{p,i}   }. \notag \\
\end{eqnarray}

Since the Wigner function does not depend on the center-of-mass coordinate of the cluster,
a Jacobi transformation~\cite{Mat97,ChenLW03,ChenLW04,LWChen06,ChenLW07} can be
performed to separate the center-of-mass coordinate and the relative coordinates of
constituent particles of the cluster, i.e.,
\begin{eqnarray}\label{EqB5}
  \left(
  \begin{array}{c}
\mathbf{R} \\
\mathbf{q}_{1} \\
\cdot \\
\cdot \\
\cdot \\
\mathbf{q}_{N-1}%
\end{array} \right )
=\hat{J}\left(
\begin{array}{c}
\mathbf{x}_{1} \\
\mathbf{x}_{2} \\
\cdot \\
\cdot \\
\cdot \\
\mathbf{x}_{N}%
\end{array}%
\right),
\left(
\begin{array}{c}
\mathbf{P} \\
\mathbf{k}_{1} \\
\cdot \\
\cdot \\
\cdot \\
\mathbf{k}_{N-1}%
\end{array} \right )
=(\hat{J}^{-1})^T\left(
\begin{array}{c}
\mathbf{p}_{1} \\
\mathbf{p}_{2} \\
\cdot \\
\cdot \\
\cdot \\
\mathbf{p}_{N}%
\end{array}%
\right), \notag \\
\end{eqnarray}
where $\mathbf{x}_{j}$~($\mathbf{p}_{j}$) is the spatial~(momentum) coordinate
of the $j$-th constituent particle and $\hat{J}$ is Jacobi matrix whose property
can be found in Appendix~\ref{AppJacobi}.
In particular, the center-of-mass position vector of the cluster $\mathbf{R}$
and the relative spatial coordinate vectors $\mathbf{q}_{i}$ can be expressed as
\begin{eqnarray}
\mathbf{R} &=& \frac{\sum_{j=1}^{N}m_{j}\mathbf{x}_{j}}{\sum_{j=1}^{N}m_{j}}, \label{EqJCM} \\
\mathbf{q}_{i} &=& \sqrt{\frac{i}{i+1}}\Bigg(\frac{\sum_{j=1}^{i}m_{j}\mathbf{x}_{j}}{\sum_{j=1}^{i}m_{j}}-\mathbf{x}_{i+1}\Bigg). \label{EqJRel}
\end{eqnarray}
Correspondingly, in the momentum space,
$\mathbf{P}$ is the total momentum of the cluster and
$\mathbf{k}_i$ are the relative momentum vectors.
It should be noted that the Jacobi matrix $\hat{J}$ we use here is different
from that in Ref.~\cite{Cho11prc}, but the physical results should be independent
of the choice of coordinate transformation.
The modulus of determinant of the Jacobi matrix
is $|\hat{J}|=N^{-\frac{1}{2}}$, and one then has the following identity
\begin{eqnarray}\label{EqB7}
  \prod_{i=1}^N\text{d}^3x_i\text{d}^3p_i&=&\text{d}^3 R\text{d}^3P\prod_{i=1}^{N-1}\text{d}^3q_i\text{d}^3k_i.
\end{eqnarray}
From the above identity, Eq.~(\ref{EqB3}) can be simplified as
\begin{eqnarray} \label{EqB4}
N_c &\approx & g_c \bigg[\prod_{i=1}^{N} N_i \bigg] \frac{V'(2\pi \mu_0T)^{\frac{3}{2}} \int \rho_c^W \prod_{i=1}^{N-1} \text{e}^{-\frac{k_{T,i}^2}{2\mu_{i} T}} \text{d}^3q_i\text{d}k^3_i }{\prod_{i=1}^{N}  V' V_{p,i}   }. \notag \\
\end{eqnarray}
where
\begin{eqnarray}
\mu_{i} = \frac{i+1}{i} \frac{m_{i+1} \sum_{k=1}^{i}m_k}{\sum_{k=1}^{i+1}m_k}, ~~~~(1\leq i\leq N-1)
\end{eqnarray}
is the reduced mass related to the relative coordinates~(see Appendix~\ref{AppJacobi}
for more details); $\mu_0=\sum_{i=1}^{N} m_i$ is the total mass of constituent particles inside the cluster,
which is equal to the rest mass of the cluster if the binding energy of the cluster is neglected;
$V'$ and $(2\pi \mu_0T)^{\frac{3}{2}}$ are the spatial and momentum
effective volumes, respectively, for the center-of-mass motion of the cluster.
To obtain Eq.~(\ref{EqB4}), following Refs.~\cite{ChenLW07,Cho11,Cho11prc},
we have assumed $\eta_i \approx y_i$, and this can be justified from the fact that the coalescence probability is
highly suppressed when the relative position or the relative momentum
of the constituent particles inside the cluster is large.
One sees that
the integration with respect to momentum in Eq.~(\ref{EqB4}) is three-dimensional,
which is different from Refs.~\cite{ChenLW07,Cho11,Cho11prc} where the momentum
integration is assumed to be two-dimensional in transverse plane.

The denominator in Eq.~(\ref{EqB4}) can be re-expressed as
\begin{eqnarray}\label{EqB9}
 \prod_{i=1}^{N}  V' V_{p,i} &=&  \prod_{i=1}^{N}  V' (2\pi T m_i)^{\frac{3}{2}}   \notag \\
                             &=& N^{-\frac{3}{2}} V'(2\pi\mu_0 T)^{\frac{3}{2}}\prod_{i=1}^{N-1}  V' (2\pi \mu_i T)^{\frac{3}{2}},
\end{eqnarray}
where the factor $N^{-\frac{3}{2}}$ comes from the coordinate
transformation and will be canceled out at last.
It should be noted that to obtain Eq.~(\ref{EqB9}), Eq.~(\ref{AppD5}) has been used.

We note that the contributions of center-of-mass coordinate in the numerator
and denominator (i.e., Eq.~(\ref{EqB9})) of Eq.~(\ref{EqB4})
cancel out, and Eq.~(\ref{EqB4}) can be written as
\begin{eqnarray}\label{EqB10}
N_c &\approx & g_c \bigg[\prod_{i=1}^{N} N_i \bigg] \frac{ \int
   \rho_c^W \prod_{i=1}^{N-1}  \text{e}^{-\frac{k_{T,i}^2}{2\mu_i T}} \text{d}^3q_i \text{d}^3k_i }{N^{-\frac{3}{2}}\prod_{i=1}^{N-1}  V'(2\pi T \mu_i)^{\frac{3}{2}}  }.
\end{eqnarray}
Furthermore, for simplicity, we assume that the Wigner function of
the cluster can be factorized into the Wigner functions of each
relative coordinate, i.e.,
$\rho_c^W = \prod_{i=1}^{N-1}\rho_{c,i}^W$,
then Eq.~(\ref{EqB10}) can be recast into
\begin{eqnarray}\label{EqB11}
N_c &\approx & g_c N^{\frac{3}{2}} \bigg[\prod_{i=1}^{N} N_i \bigg] \prod_{i=1}^{N-1} \frac{ \int
   \rho_{c,i}^W  \text{e}^{-\frac{k_{T,i}^2   }{2\mu_i T}} \text{d}^3q_i \text{d}^3k_i}{  V' (2\pi T \mu_i)^{\frac{3}{2}} }.
\end{eqnarray}
If we adopt the harmonic oscillator assumption for the potential of constituent
particles~(see Appendix~\ref{AppOsc}),
and consider a mixed ensemble with a definite orbital angular momentum
state~($l$) in the lowest energy state with $n=l$, then the spatial integration
of Wigner function can be obtained as
\begin{eqnarray}\label{EqB12}
\int \rho_{c,i}^W \text{d}^3q_i &=& \frac{(2\pi)^3}{2l+1} \sum_{m=-l}^l \bigg |\tilde{\psi}(k_i,\theta,\phi)_{nlm}\bigg |^2  \notag \\
                         &=& (2\pi)^3P(k_i),
\end{eqnarray}
where $(2\pi)^3$ comes from the normalization of Wigner function
$\int \rho_{c,i}^W \text{d}^3q_i \text{d}^3k_i = (2\pi\hbar)^3$
with the reduced Plank constant $\hbar$ taken to be unity, and the
ensemble averaged probability distribution is represented by
$P(k_i) = \frac{(4\pi\sigma_i^2)^{\frac{3}{2}}}{(2\pi)^3}\frac{(2\sigma_i^2k_i^2)^l}{(2l+1)!!}e^{-\sigma_i^2 k_i^2}$
with $\sigma_i^2 = \frac{1}{\mu_i w}$ where $w$ is the frequency of
the harmonic oscillator~(see Appendix~\ref{AppOsc}) and it can be
determined by the RMS radius~($r_{\text{rms}}$) of the cluster
(see Eq.~(\ref{AppD10}) in Appendix~\ref{AppJacobi}).
After the momentum integration, the multiplicity of the cluster can
be obtained as
\begin{eqnarray}\label{EqB13}
N_c   &\approx &g_c N^{\frac{3}{2}} \bigg[\prod_{i=1}^{N} N_i \bigg] \prod_{i=1}^{N-1} F(\sigma_i,\mu_i,l_i,T),
\end{eqnarray}
where $F(\sigma,\mu,l,T)$ is expressed as~(see Appendix~\ref{AppCoalF})
\begin{eqnarray}\label{EqB14}
&&F(\sigma,\mu,l,T) \notag  \\
                  &=&\frac{(4\pi\sigma^2)^{\frac{3}{2}}}{V'(\frac{2T}{w})^{\frac{1}{2}}(1+\frac{2T}{w})} \Big(\frac{\frac{2T}{w}}{\frac{2T}{w} +1}\Big)^l G\Big(l, (\frac{2T}{w})^{\frac{1}{2}}\Big),
\end{eqnarray}
with the function $G(l,x)$ defined as a ratio of two hypergeometric
function (see Appendix~\ref{AppAngMom}).

Finally. we obtain the following analytical coalescence formula, i.e.,
\begin{eqnarray}\label{EqB15}
 N_c
 &\approx &g_c \mu_0^{\frac{3}{2}}\bigg[\prod_{i=1}^{N} \frac{N_i}{m_i^{\frac{3}{2}}} \bigg]  \times \notag \\ &&\prod_{i=1}^{N-1}\frac{(\frac{4\pi}{w})^{\frac{3}{2}}}{V'(\frac{2T}{w} )^{\frac{1}{2}}(1+\frac{2T}{w})} \Big(\frac{\frac{2T}{w}}{\frac{2T}{w}+1}\Big)^{l_i} G\Big(l_i,(\frac{2T}{w})^{\frac{1}{2}}\Big). \notag \\
\end{eqnarray}
Since $V' = V y_L$, we can express the multiplicity per unit rapidity as
\begin{eqnarray}\label{EqB16}
 \frac{dN_c}{dy}
 &\approx &g_c \mu_0^{\frac{3}{2}}\bigg[\prod_{i=1}^{N} \frac{\frac{dN_i}{dy_i}}{m_i^{\frac{3}{2}}} \bigg] \times  \notag \\ &&\prod_{i=1}^{N-1}\frac{(\frac{4\pi}{w})^{\frac{3}{2}}}{V(\frac{2T}{w} )^{\frac{1}{2}}(1+\frac{2T}{w})} \Big(\frac{\frac{2T}{w}}{\frac{2T}{w}+1}\Big)^{l_i} G\Big(l_i,(\frac{2T}{w})^{\frac{1}{2}}\Big). \notag \\
\end{eqnarray}

\subsection{Relativistic corrections}
In the derivation of Eqs.~(\ref{EqB15}) and~(\ref{EqB16}), a non-relativistic
approximation of the momentum distribution has been adopted. However, it
should be noted that the relativistic effect could decrease the coalescence
probability as shown in the study for the exotic $\Theta^+$ production in
relativistic heavy-ion collisions~\cite{ChenLW04}. It is thus interesting
and important to include this effect in the analytical coalescence formula.
We use $g_{\text{rel}}$ to denote the relativistic correction factor.

For a cluster, it is reasonable to assume that the center-of-mass motion
should obey relativistically covariant one-particle momentum distribution,
while the relative motions of constituent particles inside the cluster
should be approximately non-relativistic because the relative momentum
and spatial coordinates have to be small, otherwise the coalescence
probability of forming a cluster is highly suppressed.
This means that the relativistic
correction of the Wigner function integration should be small, and
thus the leading order relativistic correction should mainly come from
the one-particle momentum distribution. From Eqs.~(\ref{EqA5})
and (\ref{EqA5-1}) and keeping the leading order correction in
the latter, one obtains the effective volume in momentum
space as $V_{p} = (2\pi T m)^{\frac{3}{2}} (1+\frac{15T}{8m})$.
Therefore, the relativistic correction factor $g_{\text{rel}}$ can be
expressed as
\begin{eqnarray}\label{EqRel}
g_{\text{rel}} \approx \frac{1+\frac{15}{8}\frac{T}{\mu_0}}{\prod_{i=1}^{N}(1+\frac{15}{8} \frac{T}{m_i})}.
\end{eqnarray}
On one hand, it is easy to prove that $g_{\text{rel}}$ is always less
than one and this leads to a general conclusion that the relativistic
effect tends to decrease the cluster yield.
On the other hand, one can see that when $\frac{m_i}{T}$ is much larger
than one, $g_{\text{rel}}$ approaches to unity and the relativistic
correction can then be neglected.

\subsection{Cluster finite size effects}
In evaluating the spatial part of the integration in Eq.~(\ref{EqB11}),
we have assumed that the radius $R$ of the fireball is much larger
than the size of the produced cluster.
However, for some loosely bound clusters such as $^3_\Lambda$H,
its $r_{\text{rms}}$ can be as large as $4.9$ fm~\cite{Nem00}.
It has been shown in Ref.~\cite{Sch99} that the large value of
$\frac{r_{\text{rms}}}{R}$ will decrease the value of integration.
In this work, we denote the correction factor of this cluster finite
size effect by $g_{\text{size}}$.
When $\frac{r_{\text{rms}}}{R}$ approaches to $0$, $g_{\text{size}}$ should
approach to $1$.

The cluster size correction factor $g_{\text{size}}$ cannot be analytically
obtained, but it can be calculated by a multi-dimensional
numerical integration~\cite{KJSUN15}. Based on the multi-dimensional
numerical integration, we propose the following empirical expression
for $g_{\text{size}}$ for $N$-particle~($2\le~N\le~6$)
coalescence, i.e.,
\begin{eqnarray}\label{EqSize}
g_{\text{size}}(x,N) \approx \frac{1-(2+0.6(N-1)^{0.5})x^{1.3}e^{-2x^2}}{1+(N-1)x^2},
\end{eqnarray}
where $x=\frac{r_{\text{rms}}}{R}$ is assumed to be in range [0, 0.5].
For the same $x$, we note $g_{\text{size}}$ is larger for
3-particle coalescence than for 2-particle coalescence,
but the difference is relatively small compared with their
own values. This empirical expression can be applied to central
heavy-ion collisions at both RHIC and LHC energies.
For coalescence of constituent particles with different masses, we
notice that the mass difference tends to decrease the value of
$g_{\text{size}}$, but the correction is not significant.

\section{Discussions}
Combining the relativistic correction factor $g_{\text{rel}}$ and
the cluster size correction factor $g_{\text{size}}$ with Eq.~(\ref{EqB16}),
we obtain the following full analytical coalescence formula for the
cluster multiplicity per unity rapidity, i.e.,
\begin{eqnarray}\label{COALSH}
 \frac{dN_c}{dy}
 &\approx &g_{\text{rel}} g_{\text{size}} g_c \mu_0^{\frac{3}{2}}\bigg[\prod_{i=1}^{N} \frac{\frac{dN_i}{dy_i}}{m_i^{\frac{3}{2}}} \bigg] \times  \notag \\ &&\prod_{i=1}^{N-1}\frac{(\frac{4\pi}{w})^{\frac{3}{2}}}{V(\frac{2T}{w} )^{\frac{1}{2}}(1+\frac{2T}{w})} \Big(\frac{\frac{2T}{w}}{\frac{2T}{w}+1}\Big)^{l_i} G\Big(l_i,(\frac{2T}{w})^{\frac{1}{2}}\Big). \notag \\
\end{eqnarray}
We denote the above formula as COAL-SH.
For comparison, the COAL-Ex formula (i.e., Eq.~(18) of
Ref.~\cite{Cho11prc}) reads
\begin{eqnarray}\label{EqAna1}
\frac{dN_c}{dy}
 &\approx &g_c \mu_0^{\frac{3}{2}} \bigg[\prod_{i=1}^{N} \frac{\frac{dN_i}{dy_i}}{m_i^{\frac{3}{2}}} \bigg]
        \times  \notag \\ &&\prod_{i=1}^{N-1}\frac{(\frac{4\pi}{w})^{\frac{3}{2}}}{V(1+\frac{2T}{w})} \Big(\frac{\frac{2T}{w}}{\frac{2T}{w}+1}\Big)^{l_i} \frac{(2l_i)!!}{(2l_i+1)!!}.
\end{eqnarray}

One can see that compared with the formula COAL-Ex,
besides the additional factors $g_{\text{rel}}$ and $g_{\text{size}}$,
the new formula COAL-SH displays very similar structure but different dependence
on the $w$ parameter and the orbital angular momentum quantum numbers,
which will be discussed in the following.

\subsection{Suppression from orbital angular momentum}
In the COAL-SH formula (Eq.~(\ref{COALSH})), the suppression factor
from non-zero orbital angular momentum quantum number $l$ is
\begin{eqnarray}\label{EqAnaB1}
\mathcal{S}(l,x)= \Big(\frac{x^2}{1+x^2}\Big)^lG(l,x),
\end{eqnarray}
with $x = (\frac{2T}{w})^{\frac{1}{2}}$.
From the property of $G(l,x)$~(see Appendix~\ref{AppAngMom}),
one can easily show
\begin{eqnarray}\label{EqAnaB2}
\mathcal{S}(l,x)<1,
\end{eqnarray}
indicating the suppression feature due to the finite orbital angular
momentum quantum number $l$.

For $x \gg 1$, one can show that $\mathcal{S}(l,x)$ is very close
to $1$ for all $l$. Taking $l=1$ as an illustration, one has
\begin{eqnarray}\label{EqAnaB3}
\mathcal{S}(1,x) = \frac{x^2}{1+x^2}G(1,x) =\frac{x^2+\frac{1}{3}}{x^2+1}\sim 1,
\end{eqnarray}
for $x \gg 1$.
This feature is consistent with the results in Ref.~\cite{Mul06} where the
orbital angular momentum suppression factor is shown to
approach to unity in high temperature limit (large $T$).
This is a little bit different from the result of the COAL-Ex formula
(Eq.~(\ref{EqAna1})) from which one can see the value of the suppression
factor is $\frac{2}{3}$ for $l=1$ at the limit of $x \gg 1$.

\subsection{Saturation with the cluster size}
According to Eq.~(\ref{AppD10}) in Appendix~{\ref{AppJacobi}}, the value of
harmonic oscillator frequency $w$ is inversely proportional to the $r^2_{\text{rms}}$
value of the produced cluster. Under the limit of $\frac{2T}{w} \gg 1$ for large
cluster size (but the $r_{\text{rms}}$ of the cluster is still assumed to be smaller
than the fireball radius $R$ ), the COAL-SH formula Eq.~(\ref{COALSH}) can be further
simplified to be
\begin{eqnarray}\label{EqAnaC1}
 \frac{dN_c}{dy} &\approx &g_{\text{rel}} g_{\text{size}}g_c \mu_0^{\frac{3}{2}}\bigg[\prod_{i=1}^{N} \frac{\frac{dN_i}{dy_i}}{m_i^{\frac{3}{2}}} \bigg] \prod_{i=1}^{N-1}\frac{(4\pi)^{\frac{3}{2}}}{V (2T)^{\frac{3}{2}}}.
\end{eqnarray}
The above expression indicates that the $\frac{dN_c}{dy}$ will be saturated to
a constant at the limit of large (small) $r_{\text{rms}}$~($w$).
In particular, for $R\gg r_\text{rms}$, the cluster size correction factor
$g_\text{size}$ approaches to unity and Eq.~(\ref{EqAnaC1}) becomes
\begin{eqnarray}\label{EqAnaC2}
 \frac{dN_c}{dy} &\approx &g_{\text{rel}}g_c \mu_0^{\frac{3}{2}}\bigg[\prod_{i=1}^{N} \frac{\frac{dN_i}{dy_i}}{m_i^{\frac{3}{2}}} \bigg] \prod_{i=1}^{N-1}\frac{(4\pi)^{\frac{3}{2}}}{V (2T)^{\frac{3}{2}}}
\end{eqnarray}
which means that the size of the produced cluster has no influence
on its yields in the limits $2T\gg w$ and $R\gg r_\text{rms}$.

For the COAL-Ex formula, in the limit of $\frac{2T}{w} \gg 1$ for large
cluster size, the $\frac{dN_c}{dy}$ does not saturate with the increment
(decrease) of $r_{\text{rms}}$ ($w$).
For example, in the case of two-particle coalescence, the $\frac{dN_c}{dy}$
is proportional to $r_{\text{rms}}$~\cite{Cho11,Cho11prc}. As we will see
in the following, this difference can lead to quite different predictions
for the yield ratio $^3_{\Lambda}\text{H}/^3\text{He}$ between the COAL-Ex
and COAL-SH formulae since the hypertriton $^3_{\Lambda}\text{H}$ is a very
loosely bound state with a huge radius of about $4.9$ fm~\cite{Nem00}.

\section{Applications}
In this section, we first apply COAL-SH and COAL-Ex to investigate the
strangeness population factor $S_3 = ^3_{\Lambda}$H/($^3$He$\times \Lambda/$p),
and then to study the production of some exotic hadrons,
in central Pb+Pb collisions at $\sqrt{s_{NN}}=2.76$~TeV.
The results from thermal (statistical) model are also included for comparison.

\subsection{Strangeness population factor $S_3$}
The strangeness population factor $S_3 $
was first suggested in Ref.~\cite{Arm04} and it is expected to  be a good
representation of the local correlation between baryon number and
strangeness~\cite{Zha10,Koc05}. Therefore, it may provide valuable information
of deconfinement in relativistic heavy-ion collisions.
The $S_3$ was measured
to be  $0.60\pm 0.13(\text{stat.})\pm0.21(\text{syst.})$
for central ($0$-$10\%$ centrality) Pb+Pb collisions at
$\sqrt{s_{NN}}=2.76$ TeV~\cite{ALICE15H}.
It should be noted
that while there is negligible feed-down from heavier states into
$^3_{\Lambda}$H and $^3$He, the $\Lambda$ and p are significantly influenced
by feed-down from decays of excited baryonic states. When we calculate $S_3$
within the coalescence model, the contributions to the yields of $\Lambda$ and $p$
from electromagnetic and weak decays are excluded.

Within the formula COAL-SH, $S_3$ can be expressed as
\begin{eqnarray}\label{EqAppA1}
S_3&=&\frac{((m_p+m_n+m_\Lambda)m_p)^{\frac{3}{2}}}{((m_p+m_n+m_p)m_\Lambda)^{\frac{3}{2}}}\frac{(w_{^3\text{He}}+2 T)^2}{(w_{^3_\Lambda \text{H}}+2 T)^2}\notag \\
&& \times \frac{g_{\text{rel}}(^3_\Lambda\text{H})}{g_{\text{rel}}(^3\text{He})} \frac{g_{\text{size}}(^3_\Lambda\text{H})}{g_{\text{size}}(^3\text{He})} \\
&=& 0.845 \frac{(w_{^3\text{He}}+2 T)^2}{(w_{^3_\Lambda \text{H}}+2 T)^2}  \frac{g_{\text{rel}}(^3_\Lambda\text{H})}{g_{\text{rel}}(^3\text{He})} \frac{g_{\text{size}}(^3_\Lambda\text{H})}{g_{\text{size}}(^3\text{He})}, \notag
\end{eqnarray}
where $w_{^3\text{He}} = 0.013$~GeV and $w_{^3_\Lambda\text{H}}=1.6\times 10^{-3}$~GeV
are the corresponding harmonic oscillator frequencies of $^3\text{He}$
and $^3_\Lambda\text{H}$, respectively, and their values are obtained from
the $r_{\text{rms}}$ of $^3\text{He}$~(1.76 fm~\cite{Rop09}) and
$^3_\Lambda\text{H}$~(4.9 fm~\cite{Nem00}) through Eq.~(\ref{AppD10}).
The relativistic correction of $S_3$ due to $g_{\text{rel}}$
(Eq.~(\ref{EqRel})) can be simply neglected since one has
$g_{\text{rel}}(^3\text{He}) \approx g_{\text{rel}}(^3_\Lambda\text{H})$.
Because $^3_\Lambda$H has a much larger $r_{\text{rms}}$ than $^3$He,
one has to take account of the cluster size correction factor
$g_{\text{size}}$. By neglecting the relativistic correction,
$S_3$ can be expressed as
\begin{eqnarray}\label{EqAppA2}
S_3 &\approx & 0.845 \frac{(0.013+2 T)^2}{(0.0016+2 T)^2} \frac{g_{\text{size}}(^3_\Lambda\text{H})}{g_{\text{size}}(^3\text{He})} \notag \\
 &\approx&0.845(1+\frac{0.013}{2T})^2 \frac{g_{\text{size}}(^3_\Lambda\text{H})}{g_{\text{size}}(^3\text{He})}.
\end{eqnarray}
Now, the difference for the prediction of $S_3$ between the thermal model and
the coalescence model becomes clear: the factor $0.845$ in Eq.~(\ref{EqAppA2}) corresponds to
the prediction of thermal model at LHC energy~\cite{And11}, and
the remaining parts in Eq.~(\ref{EqAppA2}) thus correspond to the corrections from
coalescence model.

In general, the freeze-out temperature $T$ lies between $0.1$ to $0.2$~GeV
in Pb+Pb collisions that we are considering here, then
$0.845(1+\frac{0.013}{2T})^2$ is in the range of $0.9\sim 0.95$, and
one can see the temperature dependence is quite weak for this part.
Compared with the thermal model, the largest correction in the coalescence model
comes from the size effect due to different sizes of $^3_\Lambda$H and $^3$He.
The radius $R$ of the fireball at freeze-out in central Pb+Pb collisions
at $\sqrt{s_{NN}}$=2.76~TeV is about $19.7$~fm~\cite{KJSUN16}, and
according to Eq.~(\ref{EqSize}), one has
$ \frac{g_{\text{size}}(^3_\Lambda\text{H})}{g_{\text{size}}(^3\text{He})}=0.60$, which leads to
$S_3=0.54\sim 0.57$, consistent with the measured value
$0.60\pm 0.13(\text{stat.})\pm0.21(\text{syst.})$. To fit the measured
central value $0.60$, one can introduce a multi-freeze-out between nucleon
and $\Lambda$ with an earlier $\Lambda$ freeze-out~\cite{KJSUN16}.

For the formula COAL-Ex, $S_3$ can be obtained as
\begin{eqnarray}\label{EqAppA3}
S_3=\frac{((m_p+m_n+m_\Lambda)m_p)^{\frac{3}{2}}}{((m_p+m_n+m_p)m_\Lambda)^{\frac{3}{2}}}\frac{(w_{^3\text{He}}+2 T)^2}{(w_{^3_\Lambda \text{H}}+2 T)^2}\frac{w_{^3\text{He}}}{w_{^3_\Lambda\text{H}}}.
\end{eqnarray}
One can see that compared with the expression~(\ref{EqAppA1}) based on
the COAL-SH formula, the expression~(\ref{EqAppA3}) from the COAL-Ex
formula does not have correction factors $g_{\text{rel}}$
and $g_{\text{size}}$ but includes an additional factor
$w_{^3\text{He}}/w_{^3_\Lambda\text{H}} = 8.1$.
As a result, $S_3$ based on COAL-Ex is larger than about $7$
and thus the COAL-Ex formula significantly overestimates the observed
$S_3$ factor value around $0.6$ for central Pb+Pb collisions at
$\sqrt{s_{NN}}=2.76$~TeV.

\subsection{Production of exotic hadrons from quark coalescence}

In the following, we focus on the production of exotic hadrons from
quark coalescence in central Pb+Pb collisions at $\sqrt{s_{NN}}=2.76$~TeV.
Three kinds of exotic hadrons with four quark
flavors (i.e., $u$, $d$, $s$, and $c$) are considered, namely,
the exotic mesons that could be in four-quark states, the exotic
baryons that could be in five-quark states, and the exotic dibaryons
that could be in six-quark or eight states.
The detailed properties,
including mass, isospin, spin, parity, orbital angular momentum
quantum number, quark configuration and decay modes,
of these exotic hadrons can be found in Table~IV of Ref.~\cite{Cho11prc}.
For the calculations with the COAL-SH formula, the cluster size
correction factor $g_{\text{size}}$ is neglected since the RMS radii of
the exotic hadrons we are considering here are
small (less than about $2$ fm) based on the harmonic oscillator
frequencies determined in the following.

For the quark coalescence model calculations, following Refs.~\cite{Cho11,Cho11prc},
the masses of $u$($d$), $s$ and $c$ constituent quarks are taken to be
$m_{u,d} = 0.3$ GeV, $m_{s} = 0.5$ GeV and $m_{c} = 1.5$ GeV.
The temperature is set to be $T = 0.154$~GeV~\cite{Baz12,Baz14}, and we note
that a small (e.g., $20\%$) variance of the temperature value does not change our
conclusion.
In order to apply the formulae COAL-SH and COAL-Ex, one also needs
the information on the harmonic oscillator frequencies~($w$ for hadrons
with only $u$($d$) quarks, $w_s$ for hadrons with strange hadrons,
and $w_c$ for charmed hadrons), the fireball volume~($V$), the number of
constituent quarks~($N_{u(d)}$ for $u(d)$ quarks, $N_s$ for $s$ quarks
and $N_c$ for $c$ quarks).

For COAL-SH, the values of frequencies $w$, $w_s$ and $w_c$ can be
deduced from the hadron size through Eq.~(\ref{AppD10}).
In particular, the values of $w$ and $w_s$ have been determined to
be $w = 0.184$~GeV and $w_s = 0.078$~GeV in Ref.~\cite{KJSUN16-1}.
For the value of $w_c$, we obtain $w_c = 0.087$ GeV by using the
value $0.546$ fm for the RMS radius of
$\Omega_{ccc}$~\cite{Zhu14}. With $w = 0.184$~GeV, $w_s = 0.078$~GeV
and $w_c = 0.087$ GeV, we obtain reasonable RMS radius values for
various hadrons, namely, $0.84$ fm for protons ($p$),
$0.87$ fm for $\phi$ mesons, $1.1$ fm for $\Xi^-$,
$1.0$ fm for $\Omega^-$, $1.2$ fm for $\Lambda$,
and $0.98$ fm for $D^0$.
Finally, the parameters $N_{u(d)}$, $N_s$ and $N_c$ can be obtained
from fitting the measured yields~($dN/dy$ at midrapidity) of $p$~\cite{Abe12},
$\Lambda$~\cite{Abe13}, $\phi$~\cite{Abe15}, $\Xi^-$~\cite{Abe14},
$\Omega^-$~\cite{Abe14}, and $D^0$~\cite{WangRQ16}.

\begin{table}[!h]
  \caption{Model parameters of COAL-SH, COAL-Ex and the thermal model,
  i.e., freeze-out temperature $T$ (GeV), fireball
  volume $V$~(10$^3$ fm$^3$), harmonic oscillator frequency $w$ (GeV), $w_s$ (GeV) and
  $w_c$ (GeV), baryon chemical potential $\mu_B$, strangeness chemical potential $\mu_S$, and
  charm fugacity $\gamma_c$, for central Pb+Pb collisions at $\sqrt{s} = 2.76$~TeV.}
  \centering
  \begin{tabular}{c|c|c|c|c|c|c|c|c}
          \hline \hline
           & $T$ & $V$ & Nu/3 & Ns/3&Nc/3 & $w$ &$w_s$&$w_c$ \\
           \hline
                      COAL-SH&0.154 & 13.8 &462 & 196 &27&0.184&0.078&0.087 \\
           \hline
           COAL-Ex &0.154 & 6.06 &267 & 156 &20&0.55&0.519&0.385\\
           \hline\hline
                & $T$ & $V$ & $\mu_B$&$\mu_S$&$\gamma_c$  \\
           \hline
           Thermal &0.153& 6.45& 0&0  &61.6  \\
          \hline \hline
  \end{tabular}
  \label{Tab4}
\end{table}

For the yields of $p$, $\Lambda$, $\phi$, $\Xi^-$, $\Omega^-$
and $D^0$ in central Pb+Pb collisions at $\sqrt{s_{NN}}=2.76$~TeV,
it should be noted that the weak decays have already been
corrected in the experimental data, but not for the strong decays
and electromagnetic decays. In order to compare with experiment
results, we have to include the contributions from strong and electromagnetic
decays in quark coalescence model.
Following Refs.~\cite{Cho11,Cho11prc,KJSUN16-1}, we assume the relations
$N_{\Lambda(1115)}^{\text{measured}} = N_{\Lambda(1115)} + \frac{1}{3} N_{\Sigma(1192)} + (0.87+\frac{0.11}{3})N_{\Sigma(1385)} = 7.44N_{\Lambda(1115)}$,
$N_{p}^{\text{measured}} = N_{p} + N_{\Delta^{++}(1232)} + \frac{1}{2}N_{\Delta^{+}(1232)} + \frac{1}{2}N_{\Delta^{0}(1232)} = 5N_p$, $N_{\Xi^-}^{\text{measured}} = N_{\Xi^-} + \frac{1}{2}N_{\Xi(1530)} = 3N_{\Xi^-}$,
and $N_{D^0}^{\text{measured}} = N_{D^0} + N_{D^{*0}} +0.677N_{D^{+*}} = 6.0N_{D^0}$.
For $\phi$ and $\Omega^-$, we assume no strong and electromagnetic
decay corrections, and thus $N_{\phi}^{\text{measured}} = N_{\phi}$ and
$N_{\Omega^-}^{\text{measured}} = N_{\Omega^-}$.
Here $N_{\Lambda(1115)}$, $N_p$, $N_{\Xi^-}$, $N_{\phi}$,
$N_{\Omega^-}$ and $N_{D^0}$ represent the corresponding hadron
multiplicity obtained directly from the quark coalescence model.

For COAL-SH, by fitting the measured yields of $p$, $\Lambda$, $\phi$,
$\Xi^-$, $\Omega^-$ and $D^0$, we obtain the fireball volume
$V = 1.38\times 10^4$ fm$^3$, the constituent quark numbers
$N_{u(d)} = 3\times 462$, $N_s = 3\times 196$ and $N_c = 3\times 27$.

For the formula COAL-Ex, we find it cannot reasonably fit the measured
yields of $p$, $\Lambda$, $\phi$, $\Xi^-$, $\Omega^-$ and $D^0$ if the
frequency parameters $w$, $w_s$ and $w_c$ are taken to have the same
values as those in the COAL-SH formula.
In Refs.~\cite{Cho11,Cho11prc}, for central Au+Au collisions at
$\sqrt{s_{NN}}=200$~GeV and central Pb+Pb collisions at
$\sqrt{s_{NN}}=5$~TeV, these frequencies were assumed to be free
parameters and determined to be $w=0.55$ GeV, $w_s=0.519$ GeV and
$w_c=0.385$ GeV by reproducing the yields of $w(782)$, $\rho(770)$,
$\Lambda(1115)$ and $\Lambda_c(2286)$ predicted from the thermal
model.
If we take these frequency values for the formula COAL-Ex,
we then obtain the fireball volume $V = 6.06\times 10^3$ fm$^3$, the
constituent quark numbers $N_{u(d)} = 3\times 267$, $N_s = 3\times 156$
and $N_c = 3\times 20$.

For the thermal model (see Refs.~\cite{Cho11,Cho11prc}
for details), the baryon and strange chemical potentials are set
to be zero. And by fitting the measured yields of $p$, $\Lambda$,
$\phi$, $\Xi^-$, $\Omega^-$ and $D^0$, we obtain the temperature
$T = 0.153$~GeV, the fireball volume $V = 6.45\times 10^3$ fm$^3$,
and the charm fugacity $\gamma_c = 61.6$.

\begin{table}[!h]
\centering
\caption{$dN/dy$ at midrapidity of $p$, $\Lambda$, $\phi$,
$\Xi^-$, $\Omega^-$ and $D^0$ in central Pb+Pb collisions at $\sqrt{s}=2.76$~TeV
predicted from COAL-SH, COAL-Ex and the thermal model. The experimental data and
the coalescence factor $g_c$ are also included.}
\centering
        \begin{tabular}{c|c|c|c|c|c|c}
          \hline \hline
          Hadron & $p$ & $\phi$ & $\Xi ^-$ & $\Omega^-$  & $\Lambda$&$D^0$ \\
          $g_c$ &  $\frac{2}{3^3\times 2^3}$ &$\frac{3}{3^2\times 2^2}$ & $\frac{2}{3^3\times 2^3}$ & $\frac{4}{3^3\times 2^3}$  &$\frac{2}{3^3\times 2^3}$ & $\frac{1}{3^2\times 2^2}$ \\
          Exp.&$34     $& $13.8        $&$3.34\        $&$0.58      $& $26     $&8.4\\
              &$  \pm 3$& $    \pm 1.77$&$    \pm 0.25 $&$    \pm0.1$& $  \pm 3$& - \\
          COAL-SH & 34.4 & 13.8&3.23 & 0.64&		26.3&8.4 \\
		  COAL-Ex & 36.2 & 13.8&3.16 & 0.71&		22.5& 8.4 \\
		  Thermal & 31.1 & 16.7&3.48 & 0.60&		19.3&8.4 \\
          \hline \hline
\end{tabular}
\label{Tab3}
\end{table}

Table~\ref{Tab4} summarizes the model parameters for COAL-SH, COAL-Ex
and the thermal model. Table~\ref{Tab3} displays the predictions from
COAL-SH, COAL-Ex and the thermal model, for
the yields of $p$, $\Lambda$, $\phi$, $\Xi^-$, $\Omega^-$ and $D^0$
in central Pb+Pb collisions at $\sqrt{s} = 2.76$~TeV.
Also included in Table \ref{Tab3} are the corresponding experimental
data and the coalescence factors $g_c$ due to spin and color degrees
of freedom.
From Table~\ref{Tab3},
one can see COAL-SH gives a best fit, COAL-Ex also gives a nice fit,
while the thermal model does not fit the $\Lambda$ yield well.
We note that COAL-SH can give a similar good description for the yields of $p$, $\Lambda$,
$\phi$, $\Xi^-$ and $\Omega^-$ as the more realistic multi-dimensional
numerical integration method, and the latter has been shown to describe very well
the spectra and yields of these hadrons~\cite{KJSUN16-1}. This implies that the COAL-SH
formula can be served as a very useful tool to evaluate the yield of clusters produced
in relativistic heavy-ion collisions.

Shown in Tab.~\ref{Tab5} are the predicted exotic hadron yields from
COAL-SH, COAL-Ex and the thermal model.
The properties of these exotic hadrons can be found in Table IV
of Refs.~\cite{Cho11prc}.
In Tab.~\ref{Tab5}, most of the predicted yields from COAL-SH and
COAL-Ex are very close and the differences are within about
factor two.
However, there are also some cases that the difference between the
COAL-SH and COAL-Ex predictions is larger than factor two. This is mainly
due to different suppressions of non-zero orbital angular momentum states
of the exotic hadrons within these two formulae.
Compared with the predictions from thermal model, most of the results
of COAL-SH are much smaller by about one order, and even smaller by about
two orders for some cases. However, when the masses of the exotic hadrons,
such as $a_0(980)$, $X(3872)$, $Z^+(4430)$ and $\Lambda(1450)$, are much
larger than the total mass of their constituent quarks,
their yields from COAL-SH are larger than those from the
thermal model.

In particular, it is suggested that the mesons $f_0(980)$, $a_0(980)$,
$D_s(2371)$ and $X(3872)$ could be either in normal two-quark states
or in exotic four-quark states.
Similarly, the baryon $\Lambda(1405)$ could be either in normal three-quark
state (i.e., $uds(L=1)$) or in exotic five-quark state (i.e., $qqqs{\bar q}$).
It is interesting to see that the predicted
yields of these hadrons in exotic four(five)-quark states are smaller by about
two orders than those in normal two(three)-quark states, indicating that
measuring the yields of these hadrons in central Pb+Pb collisions at
$\sqrt{s} = 2.76$~TeV will be potentially useful to identify the internal
quark configuration of these hadrons.

\begin{widetext}

\begin{table*}[!h]
\small
  \caption{$dN/dy$ at midrapidity for some exotic hadrons with various quark configurations
  in central Pb+Pb collisions at $\sqrt{s}=2.76$~TeV predicted from COAL-SH, COAL-Ex and
  the thermal model.}
  \centering
  \begin{tabular}{c|c|c|c|c|c}
  \hline\hline
  Models              & COAL-SH & COAL-Ex &COAL-SH & COAL-Ex & Thermal \\
         \hline
  Quark configuration & 2q/3q/6q & 2q/3q/6q  &4q/5q/8q&4q/5q/8q &  -  \\
         \hline
  Mesons:      \\ \hline
  $f_0(980)$&  35.5,3.98($s\bar{s}$)&7.24,1.14($s\bar{s}$)& 5.90$\times 10^{-2}$&  	5.54$\times 10^{-2}$&6.85 \\
  $a_0(980)$&  72.4&19.6&8.54$\times 10^{-2}$&13.7$\times 10^{-2}$&	20.6 \\
  $K(1460)$& -&-&4.03$\times10^{-1}$	&3.35$\times10^{-1}$&10.0$\times10^{-1}$\\
  $D_s(2371)$& 3.38$\times10^{-1}$&1.31$\times10^{-1}$&3.76$\times10^{-3}$&	7.52$\times10^{-3}$&2.16$\times10^{-1}$ \\
  $T_{cc}$&  -&-& 6.28$\times10^{-4}$&9.04$\times10^{-4}$&5.17$\times10^{-3}$ \\
  $X(3872)$& 5.37$\times10^{-2}$&6.34$\times10^{-3}$	&6.28$\times10^{-4}$&9.04$\times10^{-4}$&3.26$\times10^{-3}$  \\
  $Z^+(4430)$& -&-&5.35$\times10^{-4}$&2.68$\times10^{-4}$&1.01$\times10^{-4}$    \\
  \hline
  Baryons:   \\\hline
  $\Lambda(1405)$& 3.06& 0.75&4.35$\times10^{-2}$&3.46$\times10^{-2}$&	1.36    \\
  $\Theta^+$&  -&-&3.76$\times10^{-2}$&0.86$\times10^{-2}$	& 6.75$\times10^{-1}$	   \\
  $\bar{K}KN$&  -&-&2.14$\times10^{-2}$&0.55$\times10^{-2}$&1.44$\times10^{-1}$	 \\
  $\bar{D}N$&  -&-& 2.64$\times10^{-3}$&6.28$\times10^{-3}$&2.56$\times10^{-2}$	\\
  $\bar{D}^*N$& -&-&3.86$\times10^{-3}$&1.32$\times10^{-3}$&2.36$\times10^{-2}$	\\
  $\Theta_{cs}$& -&-& 1.22$\times10^{-3}$&1.12$\times10^{-3}$&1.63$\times10^{-2}$ \\
  \hline
  Dibaryons:  \\ \hline
  $H$& 6.30$\times10^{-4}$	&5.32$\times10^{-4}$&-&-&5.37$\times10^{-3}$ \\
  $\bar{K}NN$&3.91$\times10^{-3}$&8.46$\times10^{-3}$& 4.49$\times10^{-5}$&	3.01$\times10^{-5}$&5.71$\times10^{-3}$   \\
  $\Omega\Omega$& 3.49$\times10^{-6}$&	4.56$\times10^{-6}$&-&-&	1.47$\times10^{-5}$    \\
  $H_c^{++}$&  1.00$\times10^{-4}$&	3.52$\times10^{-4}$&-&-&1.10$\times10^{-3}$  \\
  $\bar{D}$NN& -&-& 2.12$\times10^{-6}$&1.18$\times10^{-6}$&8.24$\times10^{-5}$   \\
  \hline \hline
  \end{tabular}
  \label{Tab5}
\end{table*}

\end{widetext}

Furthermore, the exotic dibaryon $\bar{K}NN$ could be either in six-quark
state (i.e., $qqqqqs(L=1)$) or in eight-quark state (i.e., $qqqqqqs{\bar q}$)
with the latter (eight-quark state) having a yield smaller than that of the
former (six-quark state) by two orders. In particular, based on the COAL-SH
calculations, the yield ($dN/dy$ at midrapidity) of the exotic dibaryon
$\bar{K}NN$ is $3.91 \times 10^{-3}$ for six-quark state and
$4.49 \times 10^{-5}$ for eight-quark state. These yields are measurable
in central Pb+Pb collisions at $\sqrt{s} = 2.76$~TeV at LHC via the strong
decay $\bar{K}NN \rightarrow \Lambda N$.

The above results based on COAL-SH and COAL-Ex demonstrate that the yields
of various exotic hadrons sensitively depend on the quark configuration
and therefore the yield measurement of these hadrons is potentially useful
to determine the quark configuration of these exotic hadrons.
It should be noted that
these exotic hadrons could be in molecular states and the formulae COAL-SH
and COAL-Ex can also be used to evaluate the molecular state yield
if the freeze-out information is known for nucleons, $\Xi$, $\Omega$,
$\Xi_c$, $K$, $\bar K$, $D$, $\bar D$, $D^*$, $\bar D^*$ and $D_1$,
and this is beyond the scope of the present work and may be pursued in
future.

\section{conclusion}
Based on a blast-wave-like parametrization for phase-space configuration
of constituent particles at freeze-out in relativistic heavy-ion
collisions, we have derived a new approximate analytical formula
COAL-SH for cluster yield within the covariant coalescence model.
The new analytical coalescence formula COAL-SH improves some aspects of the
existing formulae by treating the integration of Wigner function
in the many-body coalescence process in full phase-space,
and thus has a good saturation property when the $r_{\text{rms}}$
of the produced clusters increases.
The corrections of relativistic effects and the finite size effects of
the produced cluster relative to the emission source are also
considered in the COAL-SH formula.

We have applied COAL-SH to investigate the
strangeness population factor $S_3 = ^3_{\Lambda}$H/($^3$He$\times \Lambda/$p)
from nucleon/$\Lambda$ coalescence and the exotic hadron production
from quark coalescence in central Pb+Pb collisions at
$\sqrt{s_{NN}}=2.76$ TeV.
The results have been compared with the prediction from the analytical
coalescence formula COAL-Ex derived by the ExHIC collaboration and
the thermal model.
Our results indicate that the COAL-SH formula can
reasonably describe the measured $S_3$ factor while the COAL-Ex formula
predicts a too large value of the $S_3$ factor. It is also indicated
that the cluster finite size effect is important for the $S_3$ evaluation
due to the large size of the hypernucleus $^3_{\Lambda}$H.

For the exotic hadron production in central Pb+Pb collisions at
$\sqrt{s_{NN}}=2.76$~TeV, we have determined the model parameters by
fitting the yields of $p$, $\Lambda$, $\phi$, $\Xi^-$, $\Omega^-$
and $D^0$. We have found that COAL-SH can nicely fit the experiment data
of these yields but COAL-Ex cannot reasonably fit the data if the harmonic
oscillator frequencies are determined from the RMS radii
of normal hadrons. By adjusting the values of the harmonic
oscillator frequencies, the COAL-Ex can reasonably fit the measured
yields. The thermal model can also fit the measured yields except that
the $\Lambda$ yield cannot be well reproduced.

For the yields of exotic hadrons, most of the predictions from COAL-SH and
COAL-Ex (with adjusted harmonic oscillator frequencies) are very close
to each other within about factor two.
COAL-SH and COAL-Ex may give significantly different predictions for
the exotic hadrons with non-zero angular momentum states as the two formulae
have different suppressions of non-zero angular momentum states.
Compared with the predictions from thermal model, most of the results
of COAL-SH are about one order smaller or even more
except for $a_0(980)$, $X(3872)$, $Z^+(4430)$ and $\Lambda(1450)$,
for which the hadron masses are much larger than the total mass of
their constituent quarks.

The new formula COAL-SH can be served as a useful tool to
study particle production in relativistic heavy-ion collisions.
It should be pointed out that for the COAL-SH formula, we have
assumed the size of the emission source for constituent particles
at freeze-out should be larger than that of the produced cluster,
and thus it cannot be simply applied to some large-size cluster
production in small collision system like $pp$ collision.
This interesting issue will be explored in the near future.

\begin{acknowledgments}
We are grateful to Che Ming Ko for helpful discussions.
This work was supported in part by
the Major State Basic Research Development Program (973 Program) in China under
Contract Nos. 2015CB856904 and 2013CB834405,
the National Natural Science Foundation of China under Grant Nos. 11625521, 11275125
and 11135011,
the Program for Professor of Special Appointment (Eastern Scholar) at Shanghai
Institutions of Higher Learning,
Key Laboratory for Particle Physics, Astrophysics and Cosmology, Ministry of
Education, China,
and the Science and Technology Commission of Shanghai Municipality (11DZ2260700).
\end{acknowledgments}

\appendix

\section{Quantum 3-dimension isotropic harmonic oscillators}
\label{AppOsc}
We consider a 3-dimension isotropic quantum
harmonic oscillator with the potential of $V(r) = \frac{1}{2}\mu \omega^2 r^2$,
where $\mu$ ($\omega$) is the mass (frequency) of the oscillator.
The wave function can be obtained by solving Sch$\ddot{o}$dinger equation
and the solution reads
\begin{eqnarray}
\label{AppA1}
\psi(r,\theta,\phi)_{nlm} = N_{sl}r^le^{- \frac{r^2}{2\sigma^2}}L_s^{l+\frac{1}{2}}\bigg(\frac{r^2}{\sigma^2}\bigg)Y_{lm}(\theta,\phi),
\end{eqnarray}
where $n = 2s+l$ is the principal quantum number related to the
energy $E = \hbar \omega (n+\frac{3}{2})$, $l$ is the orbital angular
momentum quantum number, $m=-l,...,l$ is the magnetic quantum number,
and $s$ is a non-negative integer representing radial excitation.
$N_{sl} = \sqrt{\sqrt{\frac{1}{\pi}}\frac{2^{s+l+2}s! \sigma^{-2l-3}}{(2s+2l+1)!!}}$
is a normalized constant with $\sigma^2 = \frac{1}{\mu w}$, and
$L_s^{l+\frac{1}{2}}(x)$ is the generalized Laguerre polynomials with
order of $s$. $Y_{lm}(\theta,\phi)$ is the spherical harmonic function
which is normalized as $\int |Y_{lm}|^2 d\Omega = 1$.
The RMS radius $r_{\text{rms}}$ can then be obtained via the following
expression
\begin{eqnarray}
\label{AppA2}
r_{\text{rms}}^2 = \int |\psi|^2 r^2 \text{d}^3r = \frac{3+2(l+2 s)}{2} \sigma^2.
\end{eqnarray}

The wave function in momentum~($k$) representation reads
\begin{eqnarray}
\label{AppA3}
\tilde{\psi}(k,\theta,\phi)_{nlm} &=& N_{sl}k^l\sigma^{2l+3}e^{- \frac{k^2\sigma^2}{2}}L_k^{l+\frac{1}{2}}(k^2\sigma^2)Y_{lm}(\theta,\phi). \notag \\
\end{eqnarray}
For an ensemble having the lowest energy with a given $l$, the
averaged probability distribution in momentum space can be
obtained as
\begin{eqnarray}
\label{AppA4}
P(k) &=&\frac{1}{2l+1}\sum_{m=-l}^l |\tilde{\psi}(k,\theta,\phi)_{llm}|^2  \notag \\
&=& \frac{(4\pi\sigma^2)^{\frac{3}{2}}}{(2\pi)^3}\frac{(2\sigma^2k^2)^l}{(2l+1)!!}e^{-\sigma^2 k^2} ,
\end{eqnarray}
and $P(k)$ is normalized as $\int P(k)\text{d}^3 k = 1$. To obtain Eq.~(\ref{AppA4}),
the following Uns$\ddot{o}$ld's theorem has been used
\begin{eqnarray}
\label{AppA5}
\frac{1}{2l+1}\sum_{m=-l}^l |Y_{lm}(\theta,\phi)|^2 = \frac{1}{4\pi}.
\end{eqnarray}

\section{Function $F(\sigma,\mu,l,T)$}
\label{AppCoalF}
The function $F(\sigma,\mu,l,T)$ is defined as
\begin{eqnarray}
\label{AppB1}
F(\sigma,\mu,l,T) &=& \frac{\int (2\pi)^3P(k)e^{-\frac{k_T^2}{2\mu T}} \text{d}^3k}{V' (2\pi\mu T)^{\frac{3}{2}}},
\end{eqnarray}
where $k_T = k\sin\theta$ is the transverse momentum.
With Eq.~(\ref{AppA4}), the $k$-dependent part can be
integrated out as follows
\begin{eqnarray}
\label{AppB2}
&&F(\sigma,\mu,l,T)=  \frac{\int (2\pi)^3P(k)e^{-\frac{k^2 \sin^2\theta}{2\mu T}}2\pi k^2\sin\theta dk d\theta}{V' (2\pi\mu T)^{\frac{3}{2}}} \notag \\
                  &=&\frac{(4\pi\sigma^2)^{\frac{3}{2} }(2\sigma^2)^l}{(2l+1)!! V' (2\pi\mu T)^{\frac{3}{2}}} \int k^{2l}e^{-k^2(\sigma^2+\frac{\sin^2\theta}{2\mu T})}2\pi k^2 \sin\theta dkd\theta  \notag \\
                  &=& \frac{(4\pi\sigma^2)^{\frac{3}{2} }(2\sigma^2)^l 2\pi \frac{1}{2} \Gamma(l+1+\frac{1}{2})}{(2l+1)!! V' (2\pi\mu T)^{3/2}} \int_{0}^\pi\frac{ \sin\theta d\theta}{(\sigma^2+\frac{\sin^2\theta}{2\mu T})^{l+\frac{3}{2}}} \notag \\
                  &=& \frac{(4\pi\sigma^2)^{\frac{3}{2} }\sigma^{2l} }{V' (2\mu T)^{3/2}} \int_{0}^{{\frac{\pi}{2}}}\frac{ \sin\theta d\theta}{(\sigma^2+\frac{\sin^2\theta}{2\mu T})^{l+\frac{3}{2}}} \notag \\
                  &=& \frac{(4\pi\sigma^2)^{\frac{3}{2} }(2\mu T\sigma^2)^{l} }{V' } \int_{0}^{{\frac{\pi}{2}}}\frac{ \sin\theta d\theta}{(2\mu T\sigma^2+\sin^2\theta )^{l+\frac{3}{2}}},
\end{eqnarray}
where the $\Gamma$-function identity
$\Gamma(n+\frac{1}{2}) = \frac{\sqrt{\pi}(2 n)!}{4^n n!}$ has been used.
With the function $G(l,x)$ (see Appendix~\ref{AppAngMom}), $F$ can finally
be obtained as
\begin{eqnarray}
\label{AppB3}
F(\sigma,\mu,l,T) &=&\frac{(4\pi\sigma^2)^{\frac{3}{2}}}{V'(2\mu T \sigma^2)^{\frac{1}{2}}(1+2\mu T\sigma^2)}     \notag \\
&& \times \Big(\frac{2\mu T\sigma^2}{2\mu T\sigma^2 +1}\Big)^l G\Big(l,(2\mu T \sigma^2)^{\frac{1}{2}}\Big).
\end{eqnarray}

\section{Orbital angular momentum factor}
\label{AppAngMom}
The orbital angular momentum factor $G(l,x)$ is an integration
which can be done analytically as follows
\begin{eqnarray}
\label{AppC1}
G(l,x) &=& x (1+x^2)^{l+1} \int_0^{\frac{\pi}{2}} \frac{\sin\theta d\theta}{(\sin^2 \theta + x^2)^{l+\frac{3}{2}}} \notag \\
        &=&\frac{_2F_1[\frac{1}{2},l+\frac{3}{2},\frac{3}{2},\frac{1}{1+x^2}]}{_2F_1[\frac{1}{2},\frac{3}{2},\frac{3}{2},\frac{1}{1+x^2}]}~,
\end{eqnarray}
where $_2F_1$ is the hypergeometric function.
$G(l,x)$ can also be expanded as
\begin{eqnarray}
\label{AppC2}
G(l,x) = \sum_{k=0}^{l}\frac{l!}{k!(l-k)!} \frac{1}{(2k+1)x^{2k}}.
\end{eqnarray}
From the series expansion, the following properties of $G(l,x)$
can be easily proved, i.e.,
\begin{eqnarray}
\label{AppC3}
G(l,x)&>&1 ,  \\
G(l,x)&<& \sum_{k=0}^{l}\frac{l!}{k!(l-k)!} \frac{1}{x^{2k}} = (1+\frac{1}{x^2})^l .
\end{eqnarray}
For $x^2 \gg 1$, $G(l,x)$ is thus very close to unity.

\section{Jacobi transformation and the RMS radius of clusters}
\label{AppJacobi}
The Jacobi matrix $\hat{J}$ for the coordinate transformation,
defined in Eqs.~(\ref{EqB5}), (\ref{EqJCM}) and (\ref{EqJRel}),
has some special properties.
The Jacobi matrix $\hat{J}$ is introduced to separate the
center-of-mass and the relative coordinates of a many-body system
since the Wigner function does not depend on the center-of-mass
coordinate and it can be expressed in terms of the relative
coordinates.

Considering $N$ independent constituent particles with mass
$m_i (i= 1, 2, 3, ..., N)$ in ($3N$-dimension) harmonic oscillators
with same frequency $w$, the total potential energy of the system
can be expressed as
\begin{eqnarray}
\label{AppD1}
      V &=& -\frac{1}{2}\sum_{i=1}^N(m_i w^2 r_i^2) = -\frac{w^2}{2} \mathbf{r}^T \hat{M} \mathbf{r} \notag \\
        &=& -\frac{w^2}{2} \mathbf{r}'^{T} (J^{-1})^T \hat{M} J^{-1} \mathbf{r}' =-\frac{w^2}{2} \mathbf{r}'^{T} \hat{M}' \mathbf{r}'  ,
\end{eqnarray}
where $\hat{M}$ is the mass matrix which is diagonal and
$\mathbf{r}'$ is the new coordinate vector.
The elements of transformation matrix $\hat{J}^{-1}$ is
\begin{eqnarray}
\label{AppD2}
\left\{
\begin{aligned}
\hat{J}^{-1}_{i,1} &=& ~~~~1~~~~~~~~~~~~~~~~~~~(i=1,...,N)  \\
\hat{J}^{-1}_{i,k} &=& ~~~\frac{m_k}{\sum_{j=1}^{k} m_j }\sqrt{\frac{k}{k-1}}~(k=2,...,N;k>i)  \\
\hat{J}^{-1}_{i,i} &=& -(1-\frac{m_i}{\sum_{j=1}^i m_j})\sqrt{\frac{i}{i-1}} ~~(i=2,...,N)  \\
\hat{J}^{-1}_{i,k} &=& ~~~~0~~~~~~~~~~~~~~~~~~~(\text{otherwise}).
\end{aligned}
\right.
\end{eqnarray}
The new mass matrix $\hat{M}'$ is also diagonal and the
reduced mass can be expressed as
\begin{eqnarray}
\label{AppD3}
\mu_{i-1} = \hat{M}'_{ii} &=&\sum_{j} \sum_{k} (\hat{J}^{-1})^T_{ij} \hat{M}_{jk}(\hat{J}^{-1})_{ki}    \notag \\
                      &=&\sum_k m_k (\hat{J}^{-1})_{ki}(\hat{J}^{-1})_{ki}~,
\end{eqnarray}
from which one can see that $\mu_0$ is the total mass and $\mu_{i}$ is
the reduced mass related to the constituent particles in the corresponding
relative coordinates, i.e.,
\begin{eqnarray}
\label{AppD4}
\left\{
\begin{aligned}
\mu_0 &=& \sum_k m_k (\hat{J}^{-1})_{k1} (\hat{J}^{-1})_{k1} = \sum_k m_k,   \\
\mu_{i}   &=& \frac{i+1}{i} \frac{m_{i+1} \sum_{k=1}^{i}m_k}{\sum_{k=1}^{i+1}m_k}. ~~~~(1\leq i\leq N-1).
\end{aligned}
\right.
\end{eqnarray}
From the above equations, the following identity can then be obtained
\begin{eqnarray}
\label{AppD5}
 \prod_{i=0}^{N-1} \mu_{i} = N\prod_{i=1}^{N}m_i .
\end{eqnarray}

After Jacobi transformation, the $N$ independent constituent
particles with mass $m_i$ $(i= 1, 2, 3, ..., N)$ in harmonic oscillators
with frequency $w$ are transferred into $N$ independent particles
with mass $\mu_0$ (total mass of the cluster) and $\mu_j$ $(j=1, 2, 3, ..., N-1)$
(the reduced mass related to the $N-1$ relative coordinates)
in harmonic oscillators with frequency $w$.
At the same time, the coordinates
$\mathbf{r}_i$ $(i= 1, 2, 3, ..., N)$ are correspondingly
transferred into the center-of-mass coordinate $\mathbf{R}$ and the
$N-1$ relative coordinates $\mathbf{q}_j$ $(j=0, 1, 2, 3, ..., N-1)$
as in Eq.~(\ref{EqB5}). The total potential energy of the system can be
re-expressed as
\begin{eqnarray}
\label{AppDAfterJacobi}
      V &=& -\frac{1}{2}\mu_0 w^2 R^2 - \frac{1}{2}\sum_{i=1}^{N-1}(\mu_i w^2 q_i^2).
\end{eqnarray}
The above expression is the basis of deriving the Wigner function in
Section~{\ref{AnaCoalFormula}}. The wave functions for harmonic oscillators
with mass $\mu_j$ $(j= 1, 2, 3, ..., N-1)$ and frequency $w$ can be found
in Appendix~\ref{AppOsc}.

Furthermore, the mean-square radius $r_{\text{rms}}^2$ can be obtain as
\begin{eqnarray}
\label{AppD6}
      r_N^2 &=&\frac{1}{N} \sum_{i=1}^N(\mathbf{r}_i - \mathbf{I}R)^T(\mathbf{r}_i-\mathbf{I}R) \notag \\
            &=& \frac{1}{N} \sum_{i=1}^N(\sum_{j=1}^N\hat{J}^{-1}_{ij}\mathbf{r}'_j - \mathbf{I}R)^T(\sum_{j=1}^N\hat{J}^{-1}_{ij}\mathbf{r}'_j-\mathbf{I}R) \notag \\
            &=&\frac{1}{N} \mathbf{r}'^T \hat{L}^T \hat{L}\mathbf{r}' =\frac{1}{N} \mathbf{r}'^T \hat{S}\mathbf{r}' ,
\end{eqnarray}
where we define $\hat{S} = \hat{L}^T \hat{L}$ with the matrix
$\hat{L}$ defined via the following relations, i.e.,
\begin{eqnarray}
\label{AppDL}
      \hat{L}_{i1}&=&0,   \notag \\
      \hat{L}_{ij}&=&(\hat{J}^{-1})_{ij}~~(j>1).
\end{eqnarray}
With Eq.~(\ref{AppD2}) and Eq.~(\ref{AppDL}), one can obtain the elements of
matrix $\hat{S}$, i.e.,
\begin{eqnarray}
\label{AppD7}
\left\{
\begin{aligned}
S_{ii}  &=& \frac{i}{i-1} \Bigg[\frac{(i-1)m_i^2}{(\sum_{j=1}^{j=i}m_j)^2} + \frac{(\sum_{j=1}^{j=i-1}m_j)^2}{(\sum_{j=1}^{j=i}m_j)^2}\Bigg] (i\geq 2)    \\
S_{ij} &=& 0~~~~~~~~~~~~~~~~~(i\neq j;i=j=1),
\end{aligned}
\right.
\end{eqnarray}
and then one has
\begin{eqnarray}
\label{AppD8}
\langle r_N^2 \rangle &=& \frac{1}{N}\sum_{i=1}^{N}S_{ii} \langle \mathbf{r}'_{i}\mathbf{r}'_{i} \rangle \notag \\
        &=& \frac{1}{N}\sum_{i=1}^{N-1}S_{i+1,i+1} \langle q_i^2 \rangle .
\end{eqnarray}
Furthermore, from Eq.~(\ref{AppA2}), one has
\begin{eqnarray}
\label{AppD8-1}
   \langle q_i^2 \rangle = \frac{3+2(l_i+2 s_i)}{2} \sigma_i^2,
\end{eqnarray}
where $\sigma_i^2 = 1/(\omega \mu_i)$, and $l_i$ and $s_i$ are, respectively,
the quantum numbers of orbital angular momentum and radial excitation for the
$i$-th relative coordinate.

In particular, for $l=s=0$, we have
\begin{eqnarray}
\label{AppD9}
\langle q_i^2 \rangle = \frac{3}{2} \sigma_i^2 = \frac{3}{2}\frac{1}{w\mu_i}.
\end{eqnarray}
From Eqs.~(\ref{AppD4}),~(\ref{AppD7}),~(\ref{AppD8}) and~(\ref{AppD9}),
one can then obtain the analytical formula for mean-square radius of the
cluster as
\begin{eqnarray}
\label{AppD10}
\langle r_N^2 \rangle &=& \frac{3}{2Nw}\sum_{i=1}^{N-1} \frac{1}{\mu_i}S_{i+1,i+1} \notag \\
&=& \frac{3}{2Nw}\sum_{i=2}^{N}  \frac{\frac{i}{i-1} [\frac{(i-1)m_i^2}{(\sum_{j=1}^{j=i}m_j)^2} + \frac{(\sum_{j=1}^{j=i-1}m_j)^2}{(\sum_{j=1}^{j=i}m_j)^2}] }{\frac{i}{i-1} \frac{m_i \sum_{k=1}^{i-1}m_k}{\sum_{k=1}^{k=i}m_k}}   \notag \\
      &=& \frac{3}{2Nw}\Bigg[\bigg(\sum_{i=1}^N \frac{1}{m_i}\bigg) - \frac{N}{\sum_{i=1}^N m_i}\Bigg].
\end{eqnarray}


\begin{thebibliography}{99}

\bibitem{But61} S. T. Butler and C. A. Pearson, Phys. Rev. Lett. \textbf{7}, 69 (1961).
\bibitem{Sat81} H. Sato and K. Yazaki, Phys. Lett. \textbf{B98}, 153 (1981).
\bibitem{Cse86} L. P. Csernai and J. I. Kapusta, Phys. Rep. \textbf{131}, 223 (1986).
\bibitem{Dov91} C. B. Dover, U. Heinz, E. Schnedermann, and J. Zimanyi, Phys. Rev. C \textbf{44}, 1636 (1991).

\bibitem{Gyu83} M. Gyulassy, K. Frankel, E. A. Relmer, Nucl. Phys. \textbf{A402}, 596 (1983).
\bibitem{Mat97} R. Mattiello, H. Sorge, H. St\"{o}cker, and W. Greiner, Phys. Rev. C \textbf{55}, 1443 (1997).
\bibitem{ChenLW03} L. W. Chen, C. M. Ko, and B. A. Li, Phys. Rev. C \textbf{68}, 017601 (2003);
Nucl. Phys. \textbf{A729}, 809 (2003); Phys. Rev. C \textbf{69}, 054606 (2004).
\bibitem{Oh09} Y. Oh, Z. W. Lin, and C. M. Ko, Phys. Rev. C \textbf{80}, 064902 (2009).
\bibitem{Ste12} J. Steinheimer, K. Gudima, A. Botvina, I. Mishustin, M. Bleicher, and H. St\"{o}cker, Phys. Lett. \textbf{B714}, 85 (2012).
\bibitem{ChenG12} G. Chen et al., Phys. Rev. C \textbf{86}, 054910 (2012).
\bibitem{ChenG13} G. Chen et al., Phys. Rev. C \textbf{88}, 034908 (2013).
\bibitem{Zhu15} L. L. Zhu, C. M. Ko, and X. J. Yin, Phys. Rev. C \textbf{92}, 064911 (2015).
\bibitem{Zha10} S. Zhang, J. H. Chen, H. Crawford, D. Keane, Y. G. Ma, and Z. B. Xu, Phys. Lett. \textbf{B684}, 224 (2010).
\bibitem{Xue15} L. Xue, Y. G. Ma, J. H. Chen, and S. Zhang, Phys. Rev. C \textbf{85}, 064912 (2012); Phys. Rev. C \textbf{92}, 059901 (2015).
\bibitem{KJSUN15} K. J. Sun and L. W. Chen, Phys. Lett. \textbf{B751}, 272 (2015).
\bibitem{KJSUN16} K. J. Sun and L. W. Chen, Phys. Rev. C \textbf{93}, 064909 (2016).
\bibitem{KJSUN16-1} K. J. Sun and L. W. Chen, Phys. Rev. C \textbf{94}, 064908 (2016).

\bibitem{Lin02} Z. W. Lin and C. M. Ko, Phys. Rev. Lett. \textbf{89}, 202302 (2002).
\bibitem{Vol03} S. A. Voloshin, Nucl. Phys. \textbf{A715}, 379 (2003).
\bibitem{Hwa03} R. C. Hwa and C. B. Yang, Phys. Rev. C \textbf{67}, 064902
(2003).  
\bibitem{Gre03} V. Greco, C. M. Ko, and P. L\'{e}vai, Phys. Rev. Lett. \textbf{90},
202302 (2003); Phys. Rev. C \textbf{68}, 034904 (2003). 
\bibitem{Fri03} R. J. Fries, B. M\"{u}ller, C. Nonaka, and S. A. Bass, Phys. Rev. Lett. \textbf{90}, 202303 (2003); Phys.
Rev. C \textbf{68}, 044902 (2003).  
\bibitem{Mol03} D. Molnar and S. A. Voloshin, Phys. Rev. Lett. \textbf{91}, 092301 (2003).
\bibitem{Pra05} S. Pratt and S. Pal, Phys. Rev. C \textbf{71}, 014905 (2005).
\bibitem{Sha05} F. L. Shao, Q. B. Xie, and Q. Wang, Phys. Rev. C \textbf{71}, 044903 (2005).
\bibitem{LWChen06} L. W. Chen and C. M. Ko, Phys. Rev. C \textbf{73}, 044903
(2006).  
\bibitem{Ko14} C. M. Ko, T. Song, F. Li, V. Greco, and S. Plumari, Nucl. Phys. \textbf{A928}, 234 (2014).
\bibitem{Fri08} R. Fries, V. Greco, and P. Sorensen, Annu. Rev. Nucl. Part. Sci. \textbf{58}, 177 (2008).

\bibitem{Adl03}  S. S. Adler et al., (PHENIX Collaboration), Phys. Rev. Lett. \textbf{91}, 182301 (2003).
\bibitem{Ada04}  J. Adams et al., (STAR Collaboration), Phys. Rev. Lett. \textbf{92}, 052302 (2004).

\bibitem{ChenLW04} L. W. Chen, V. Greco, C. M. Ko, S. H. Lee, and W. Liu, Phys. Lett. \textbf{B601}, 34 (2004).
\bibitem{ChenLW07} L. W. Chen, C. M. Ko, W. Liu, and M. Nielsen, Phys. Rev. C \textbf{76}, 014906 (2007).
\bibitem{Cho11}  S. Cho et al., (ExHIC Collaboration), Phys. Rev. Lett. \textbf{106}, 212001 (2011).
\bibitem{Cho11prc}  S. Cho et al., (ExHIC Collaboration), Phys. Rev. C \textbf{84}, 064910 (2011).




\bibitem{Bjo83} J. D. Bjorken, Phys. Rev. D \textbf{27}, 140 (1983).



\bibitem{Ret04} F. Reti\'{e}re and M. A. Lisa, Phys. Rev. C \textbf{70}, 044907 (2004).

\bibitem{Coo74} F. Cooper and G. Frye, Phys. Rev. D \textbf{10}, 186 (1974).   

\bibitem{Nem00} H. Nemura, Y. Suzuki, Y. Fujiwara, and C. Nakamoto, Prog. Theor. Phys. \textbf{103}, 929 (2000).

\bibitem{Sch99} R. Scheibl and U. Heinz, Phys. Rev. C \textbf{59}, 1585 (1999).

\bibitem{Mul06}   Y. Kanada-En'yo and B. M\"{u}ller, Phys. Rev. C \textbf{74}, 061901 (2006).

\bibitem{Arm04} T. A. Armstrong et al., (E864 Collaboration), Phys. Rev. C \textbf{70}, 024902 (2004).

\bibitem{Koc05} V. Koch, A. Majumder, and J. Randrup, Phys. Rev. Lett. \textbf{95}, 182301 (2005).

\bibitem{ALICE15H} J. Adam et al., (ALICE Collaboration), Phys. Lett. \textbf{B754}, 360 (2016).
\bibitem{Rop09} G. R\"{o}pke, Phys. Rev. C \textbf{79}, 014002 (2009).

\bibitem{And11} A. Andronic et al., Phys. Lett. \textbf{B697}, 203 (2011).

\bibitem{Baz12} A. Bazavov et al., Phys. Rev. D \textbf{85}, 054503 (2012).
\bibitem{Baz14} A. Bazavov et al., Phys. Rev. D \textbf{90}, 094503 (2014).

\bibitem{Zhu14} H. He, Y. P. Liu, P. F. Zhuang, Phys. Lett. \textbf{B746}, 59 (2015).

\bibitem{Abe12} B. Abelev et al., (ALICE Collaboration),  Phys. Rev. Lett. \textbf{109}, 252301 (2012).
\bibitem{Abe13} B. Abelev et al., (ALICE Collaboration),  Phys. Rev. Lett. \textbf{111}, 222301 (2013).
\bibitem{Abe15} B. Abelev et al., (ALICE Collaboration),  Phys. Rev. C \textbf{91}, 024609 (2015).
\bibitem{Abe14} B. Abelev et al., (ALICE Collaboration),  Phys. Let. \textbf{B728}, 216 (2014).
\bibitem{WangRQ16} R. Q. Wang et al., Phys. Rev. C \textbf{94}, 044913 (2016).

\end{thebibliography}
\end{document}